\documentclass[lettersize,journal]{IEEEtran}
\usepackage{amsmath,amsfonts}
\usepackage{algorithmic}
\usepackage{array}
\usepackage[caption=false,font=normalsize,labelfont=sf,textfont=sf]{subfig}
\usepackage{textcomp}
\usepackage{stfloats}
\usepackage{url}
\usepackage{cite}
\usepackage{verbatim}
\usepackage{graphicx}
\usepackage[colorlinks=true,linkcolor=blue,urlcolor=blue,citecolor=blue]{hyperref}

\usepackage[colorinlistoftodos]{todonotes}

\def\BibTeX{{\rm B\kern-.05em{\sc i\kern-.025em b}\kern-.08em
    T\kern-.1667em\lower.7ex\hbox{E}\kern-.125emX}}
\usepackage{balance}
\usepackage[export]{adjustbox}
\begin{document}
\title{\bf A Self-Healing Magnetic-Array-Type Current Sensor with Data-Driven Identification of Abnormal Magnetic Measurement Units}
\author{Xiaohu Liu, Kang Ma, Jian Liu, Wei Zhao, Lisha Peng, {\it Member, IEEE}, \\ Songling Huang, {\it Senior Member,  IEEE}, Shisong Li, {\it Senior Member, IEEE} 
\thanks{The authors are with the Department of Electrical Engineering, Tsinghua University, Beijing 100084, China.}
\thanks{Email: shisongli@tsinghua.edu.cn}}

\markboth{}{}

\maketitle

\begin{abstract}

Magnetic-array-type current sensors have garnered increasing popularity owing to their notable advantages, including broadband functionality, a large dynamic range, cost-effectiveness, and compact dimensions. However, the susceptibility of the measurement error of one or more magnetic measurement units (MMUs) within the current sensor to drift significantly from the nominal value due to environmental factors poses a potential threat to the measurement accuracy of the current sensor. In light of the need to ensure sustained measurement accuracy over the long term, this paper proposes an innovative self-healing approach rooted in cyber-physics correlation. This approach aims to identify MMUs exhibiting abnormal measurement errors, allowing for the exclusive utilization of the remaining unaffected MMUs in the current measurement process. To achieve this, principal component analysis (PCA) is employed to discern the primary component, arising from fluctuations of the measured current, from the residual component, attributed to the drift in measurement error. This analysis is conducted by scrutinizing the measured data obtained from the MMUs. Subsequently, the squared prediction error (SPE) statistic (also called $Q$ statistic) is deployed to individually identify any MMU displaying abnormal behavior. The experimental results demonstrate the successful online identification of abnormal MMUs without the need for a standard magnetic field sensor. By eliminating the contributions from the identified abnormal MMUs, the accuracy of the current measurement is effectively preserved.
\end{abstract}


\section{Introduction}

\IEEEPARstart{T}{he precise} measurement of electrical current is fundamentally crucial for a myriad of applications, including electrical energy metrology\cite{li2020research}, research on power electronic devices\cite{sun2021ltcc}, monitoring and control of power system status\cite{wu2023improved}, etc. Various current measurement techniques have been developed to date~\cite{ziegler2009current}, categorized into four main types based on their sensing principles: Faraday's law of electromagnetic induction [e.g., current transformer (CT), Rogowski coil, and zero-flux current sensor], Ohm's law (e.g., shunt resistor), Faraday effect (e.g., optic-fiber current sensor), and magnetic-field sensing (e.g., Hall current sensor).

Devices developed using these diverse current sensing principles exhibit unique advantages but also present technical challenges for further enhancement of their performance. For instance, a shunt resistor enables AC and DC measurement, but the need to pass into the current-carrying conductor limits its ease of use\cite{kyriazis2023modeling}. 
CTs offer high accuracy and stability in current measurement, leveraging ferromagnetic cores, and are extensively employed in power systems for AC metrology \cite{locci2000digital}. However, their frequency band is constrained (typically up to 1\,kHz) due to the hysteresis and eddy current effects of the ferromagnetic core. On the other hand, optic-fiber current sensors boast excellent dynamic response characteristics (up to MHz) and can measure both DC and AC currents. Nevertheless, addressing challenges related to environmental factors, such as temperature and vibration, to maintain measurement accuracy and stability remains a key hurdle in optimizing their performance \cite{jiang2023fiber}. Thus, continual exploration of novel current sensing techniques is imperative to fulfill the growing demand for accurate current measurement. 

Due to the significant advancement of the magnetic field sensing techniques (e.g., Hall effect, TMR effect, and flux-gate effect)\cite{khan2021magnetic}, magnetic-array-type current sensor \cite{moghe2012novel,li2022contactless,liu2022coreless}, based on the principle of magnetic field sensing, is recognized as having great potential for panoramic perception in smart grid, characterized with the superior performance of wide bandwidth, high linearity, large dynamic range, compact dimensions, and low cost. {Magnetic-array-type current sensors have been applied to a variety of scenarios including weak current measurement \cite{liu2023enhanced}, power measurement of residential homes \cite{liu2022semi}, condition monitoring \cite{khawaja2017estimating} or fault detection \cite{kazim2019fault} of overhead transmission lines, and energization-status identification of underground power cables \cite{zhu2017energization}.} With the increasing share of renewable energy in smart grids, the applications for magnetic-array-type current sensors are expanding.  

The most essential step before applying magnetic-array-type current sensor is to accurately determine the coefficients characterizing the mathematical relationship between the magnetic fields measured by multiple MMUs and the current to be measured. These coefficients can be calculated using the known geometry parameters of the conductor system \cite{ibrahim2020design} or obtained experimentally by applying a standard current to the conductor offline \cite{lee2018superpositioning}. {However, the offline calibration method suffers from measurement error drift due to the conductor position shift\cite{ma2019impact}. The online calibration method is much more preferred due to its simpler procedure. The coefficients are directly calculated by analytically \cite{zhang2019current} or numerically \cite{chen2022intelligent,zhu2022event} solving inverse models with the geometry parameters of the conductor system as variables and re-calibrating can be easily done if the position changes.} 

After the coefficients determination of the magnetic-array-type current senor, current can be continuously measured by multiplying the measured magnetic fields by all the MMUs with the calibrated coefficients. Thus, the long-term current measurement accuracy is highly dependent on that of each MMU. As we all know, long-term stability of measurement accuracy is a key parameter for evaluating the performance of measurement instruments. { However, when affected by environmental factors, due to the variability of the measurement performance of the MMUs, the measurement errors of a minority of MMUs may drift significantly, resulting in the degradation of the overall measurement accuracy of the magnetic-array-type current sensor.} Thus, it is extremely necessary to accurately and reliably evaluate the measurement error status of each MMU. 

The only available method for evaluating the measurement error status of each MMU is the offline calibration method, which involves detaching the magnetic-array-type current sensor from the conductor and evaluating the measurement error of each MMU with a high-precision magnetometer\cite{liu2021nonintrusive}. However, this offline method is time-consuming, labor-intensive, and involves a heavy workload.
A potential online measurement error status monitoring method is to embed a standard magnetic field generating circuit (e.g., an onboard coil) at the exact location of each MMU, allowing the specific MMU to be calibrated with the generated standard magnetic field. This online method, however, suffers from geometric dimension drift of the standard magnetic field generating circuit due to environmental factors, e.g., temperature. Currently, no studies have been reported on this online method.

With the large-scale application of distributed sensors in smart grids \cite{huang2014magnetics}, evaluating their error characteristics employing the two above-presented methods would be costly in terms of material and financial resources. Exploration of new approaches is urgent to substantially guarantee the long-term measurement stability of magnetic-array-type current sensors. Data-driven approaches, capable of recognizing the abnormal change of measurement error status of precision instruments by analyzing the measurement data of a cluster of instruments, have presented huge potential in the online detection of abnormal measurement errors of smart meters \cite{duan2023operational}, capacitor voltage transformers \cite{zhang2022online}, and fuel meters \cite{chu2023automatic}. Motivated by the successful applications of the data-driven approaches, the problem of evaluating the measurement error status of magnetic-array-type current sensors is possible to be fundamentally solved. Besides, to the best of our knowledge, in the research area of the magnetic-array-type current sensor, no relevant research has been reported yet concerning the problem of measurement error drift of MMUs through a data-driven approach. 

Based on the above analysis, this paper explores a self-healing approach, based on a data-driven approach rooted in cyber-physics correlation, to achieve online identification of abnormal MMUs in a magnetic-array-type current sensor. By eliminating the identified abnormal units exceeding their normal measurement errors, the measurement stability of the magnetic-array-type current sensor can be preserved. The proposed idea in this paper is hopeful to be applied to enhance the measurement stability and reliability of the distribution sensors for non-contact measurement of electrical parameters (e.g., current and voltage), thus facilitating the enhancement of the perception ability of the smart grid. 

The remainder of this paper is organized as follows. Section \ref{sec:02} introduces the abnormal MMU identification method. Section \ref{sec:03} presents the experimental study aimed at demonstrating the validation of the proposed data-driven approach rooted in cyber-physics correlation. Finally, the conclusion is given in section \ref{sec:04}.

\section{Method for Abnormal Magnetic Measurement Units Identification}
\label{sec:02}
\subsection{Magnetic-Array-Type Current Sensor}

\begin{figure}[!t]
\centering
\includegraphics[width=0.7\columnwidth]{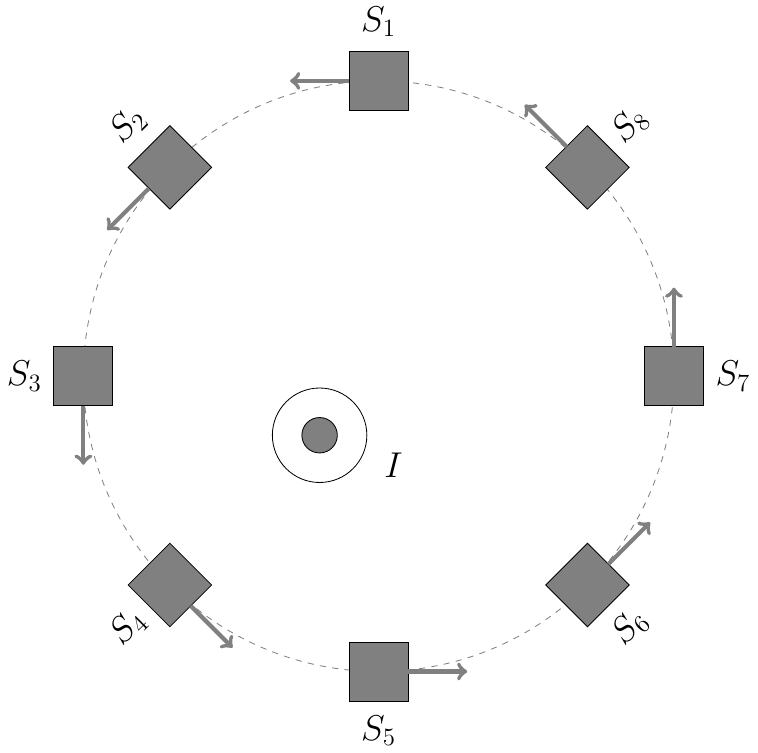}
\caption{Physical topology of the magnetic-array-type current sensor designed for non-contact current measurement. The central circle represents the current flowing through the conductor under measurement. The MMUs are depicted as gray blocks, denoted as $S_1$ through $S_N$ (in this instance, $N=8$). The directional arrow indicates the magnetically sensitive orientation of each MMU.} 
\label{fig_sa}
\end{figure}

Fig. \ref{fig_sa} presents one typical physical configuration of a magnetic-array-type current sensor for current measurement. Multiple MMUs are arranged uniformly in a circular pattern\cite{itzke2018influence}, strategically placed to measure the magnetic fields generated by a current-carrying conductor. The sensing directions of these MMUs align approximately with the tangent direction of the circle. The magnetic-array-type current sensor is proficient in gauging electrical currents across a broad spectrum, spanning from DC to wide-band AC. To enhance its efficacy in AC measurements, the implementation of the phase lock-in technique enables the sensor to accurately monitor both the amplitude and phase of a singular frequency ($f$) component. In applications such as power systems, specifically with a chosen frequency of $50$\,Hz or $60$\,Hz, this magnetic-array-type current sensor exhibits considerable potential to replace conventional CTs. This transition brings forth notable advantages in terms of flexibility, cost reduction, and compactness.

It is imperative to note that the subsequent discussion centers around the illustration of single-frequency current measurements. Utilizing the phasor representation, we articulate the operational principles and performance attributes of the magnetic-array-type current sensor. However, it is pertinent to acknowledge that the analytical framework presented herein can be extrapolated to encompass more comprehensive scenarios. In instances where the measured current comprises multiple-frequency components, such as harmonics, the analysis remains applicable and can be extended to accommodate these intricate scenarios. Without losing the generality, the current in the conductor can be measured using the following equation\cite{yu2018circular},
\begin{equation}\label{eq:1}
\dot{I}_m = \frac{1}{N}\sum_{k=1}^{N} \gamma_k\dot{B}_m^k,
\end{equation}
where $\dot{I}_m$ denotes the measured current phasor by the magnetic-array-type current sensor, $\dot{B}_m^k$ represents the magnetic field measured by the $k$th MMU in a phasor form at the fundamental frequency, $N$ is the total number of MMUs, and $\gamma_k$ is the scale factor characterizing the mathematical relationship between the magnetic field $\dot{B}_m^k$ and the current phasor $\dot{I}_m$.

After the installation of the magnetic-array-type current sensor and the calibration of the scale factor $\gamma_k$ as per (\ref{eq:1}), current measurement becomes achievable through the utilization of the magnetic fields measured by the MMUs. The accuracy of the current measurement, as indicated in (\ref{eq:1}), is closely tied to the precision of each MMU. Should there be a drift in the measurement error of one or more MMUs, the overall accuracy of the current measurement may degrade. Periodic offline calibration of all MMUs is a time-consuming and intricate process.

\begin{figure}[!t]
\centering
\includegraphics[width=0.9\columnwidth]{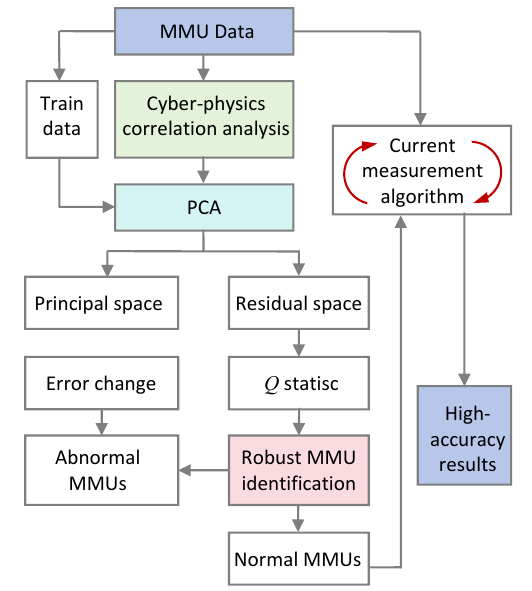}
\caption{Diagram flow of the proposed data-driven self-healing approach. } 
\label{fig:flow}
\end{figure}

In the pursuit of ensuring prolonged measurement accuracy of the magnetic-array-type current sensor, leveraging the redundancy inherent in the group of the MMUs—where the majority of MMUs operate normally—this paper introduces a data-driven self-healing approach. This approach aims to identify MMUs manifesting abnormal measurement errors, which are then eliminated. Only the remaining normal MMUs are utilized for current measurement through (\ref{eq:1}). The proposed data-driven self-healing approach is visually depicted in Fig. \ref{fig:flow}, providing an overview of its structure. Elaboration on the intricacies of the method is presented in the subsequent subsections.   

\subsection{Cyber-Physics Correlation Analysis}

In Fig. \ref{fig_sa}, the dynamic variations in the theoretical magnetic fields measured by the MMUs are inherently coupled with the fluctuations in the current. This intrinsic relationship suggests a linear correlation among the theoretical magnetic fields measured by the MMUs. The connection between the theoretical magnetic fields measured by the MMUs and the current to be measured is expressed by the following set of equations,
\begin{subequations}
\label{eq:2}
\begin{align}
c_1B_t^1&=\cdots=c_kB_t^k=\cdots=c_NB_t^N=I,\label{eq:2A}\\
\angle{\dot{B}_t^1}&=\cdots=\angle{\dot{B}_t^k}=\cdots=\angle{\dot{B}_t^N}=\angle{\dot{I}},\label{eq:2B}
\end{align}
\end{subequations}
where $\dot{B}_t^k$ and $\dot{I}$ represent the theoretical magnetic field measured by the $k$th MMU and the current phasor to be measured. Their amplitudes and phases are denoted by $B_t^k$, $\angle{\dot{B}_t^k}$, $I$, and $\angle{\dot{I}}$, respectively, while $c_k$ signifies the scale factor relating the amplitudes of $\dot{B}_t^k$ and $\dot{I}$. {Note that $c_k$ is different from $\gamma_k$ defined in (\ref{eq:1}), where $\gamma_k$ is experimentally calibrated and $c_k$ is the theoretical value of $\gamma_k$, depending on the geometry of the configured MMUs.}

The linearity in both amplitude and phase of the theoretical magnetic fields measured by the MMUs is evident from (\ref{eq:2}). The magnetic field practically measured by the $k$th MMU can be expressed as
\begin{subequations}\label{eq:3}
\begin{align}
B_m^k &=(1+\epsilon_k +\Delta{\epsilon_k})B_t^k,
\label{eq:3A}\\
\angle{\dot{B}_m^k} &= \angle{\dot{B}_t^k}+\delta_k+\Delta{\delta_k},
\label{eq:3B}
\end{align}
\end{subequations}
where $\epsilon_k$ and $\delta_k$ represent the initial measurement error for amplitude and phase, respectively, obtained through calibration and maintained as constants. $\Delta{\epsilon_k}$ and $\Delta{\delta_k}$ signify the drift in the measurement error for amplitude and phase, respectively, which may randomly vary due to environmental factors.

As elucidated by (\ref{eq:3}), the measured result by each MMU comprises two components. The first part represents the measured result with a constant measurement error, i.e., $(1+\epsilon_k)B_t^k$ or $\angle \dot{B}_t^k+\delta_k$. The second part is attributed to the drift error, denoted by $\Delta{\epsilon_k}B_t^k$ or $\Delta{\delta_k}$. Among all the MMUs, the first parts reflect the current fluctuations and are linearly related, while the second parts, caused by the random drift of measurement errors, are independent. Utilizing an effective linear correlation analysis method allows the extraction of the second part from the measured results of the MMUs, facilitating the identification of abnormal MMUs.

\subsection{Principal Component Analysis}

The principal component analysis, PCA, is a widely employed technique for dimensionality reduction, aiming to transform a large set of linearly correlated variables into a smaller, independent set\cite{greenacre2022principal}. It is particularly well-suited for separating the main component, influenced by the fluctuation of the current, from the residual component, attributed to the drift in measurement error. To achieve this, the measured results during a period when all MMUs operate under normal conditions—signifying the absence of measurement error drift—are gathered as training data for PCA model learning. These acquired data are structured as a matrix denoted by ${\bf{S}}_{L\times{N}}$, where $L$ represents the data length of each MMU, and $N$ is the number of MMUs. The training data undergo normalization using the following expression
\begin{align}\label{eq:4}
{\overline{\bf{S}}} &= ({\bf{S}}-{\bf{1}}_{L\times{1}}{\boldsymbol{\mu}}_{1\times{N}}){\bf{\Sigma}}^{-1},
\end{align}
where ${\bf{1}}_{L\times{1}}$ is a column vector with all elements as 1 and a length of $L$, ${\boldsymbol{\mu}}_{1\times{N}}$ and ${\bf{\Sigma}}$, respectively, represents a row vector and a diagonal matrix with their every element as the mean value and standard deviation of each column vector of matrix $\bf{S}$.

Let $\bf{Z}$ represent the covariance matrix of $\overline{\bf{S}}$, and singular value decomposition of $\bf{Z}$ is performed as follow
\begin{align}\label{eq:5}
{\bf{Z}} &= {{\bf{U}}_{N\times{N}}}{{\bf{\Lambda}}_{N\times{N}}}{{\bf{V}}_{N\times{N}}},
\end{align}
\noindent where $\bf{U}$ and $\bf{V}$, respectively, represents the left and right matrix, $\bf{\Lambda}$ denotes the eigenvalue matrix with its $i$th diagonal elements represented by $\sigma_i$ and all the off-diagonal elements equal to zero.  

To determine the number $m$ of the principal elements, the following criterion is commonly used
\begin{align}\label{eq:6}
\frac{\displaystyle\sum_{i=1}^m {\sigma_i^2}}{\displaystyle\sum_{i=1}^N {\sigma_i^2}}>\kappa,
\end{align}
where $\kappa$ is the threshold for determining the number of principal elements and is selected for example 0.85 empirically.  

After the number $m$ of principal elements is determined, then the principal component subspace $\bf{P}$ is established by the first $m$ column vectors of matrix $\bf{V}$ and the residual subspace $\bf{R}$ is built using the column vectors starting from the ($m$+1)th column to the last column of the matrix $\bf{V}$.

Once the principal subspace $\bf{P}$ and the residual subspace $\bf{R}$ are obtained, separation of the main component, related to the fluctuation of the current and the residual component, corresponding to the drift in measurement error, can be performed. In the long-term operation of the magnetic-array-type current sensor, the measurement results of all the MMUs are obtained as test data and are denoted by matrix $\bf{X}$, which is normalized as follows
\begin{align}\label{eq:7}
{{\overline{\bf{X}}}_{H\times{N}}} &= ({{\bf{X}}_{H\times{N}}}-{\bf{1}}_{H\times{1}}{\boldsymbol{\mu}}_{1\times{N}}){\bf{\Sigma}}^{-1},
\end{align}
where \emph{H} is the data length of the test data.

Then the normalized test data $\bf{\overline{X}}$ can be separated as the main component $\bf{\overline{M}}$ and residual component $\bf{\overline{E}}$ as follow
\begin{align}\label{eq:8}
{\bf{\overline{X}}} &= {\bf{\overline{M}}}+{\bf{\overline{E}}}.
\end{align}

In (\ref{eq:8}), $\bf{\overline{M}}$ and $\bf{\overline{E}}$ are respectively given by 
\begin{subequations}\label{eq:9}
\begin{align}
{\bf{\overline{M}}} &= {\bf{\overline{X}}}{\bf{P}}{\bf{P}}^{\rm T},\label{eq:9A}\\
{\bf{\overline{E}}} &= {\bf{\overline{X}}}{\bf{R}}{\bf{R}}^{\rm T}.\label{eq:9B}
\end{align}
\end{subequations}

For PCA theory, Hotelling $T^2$ and SPE statistic (also called $Q$ statistic) are respectively used to evaluate whether any dynamic deviation occurs in the principal component and the residual component. Because the variation of the measurement errors of the MMUs are projected to the residual component, thus the occurrence of the drift in measurement error can be sensitively detected by the change of the value of the $Q$ statistic, which is defined as follows
\begin{align}\label{eq:10}
Q_i = \sum_{j=1}^N (e_{i,j})^2,
\end{align}
where $Q_i$ represents the $Q$ statistic at the $i$th point of the test data, $j$ is used to index the MMU, $e_{i,j}$ denotes the element in the $i$th row and $j$th column of the matrix $\bf{\overline{E}}$.

A control threshold $Q_\alpha$ for the $Q$ statistic is defined as follows
\begin{align}\label{eq:11}
{Q_\alpha^2} &= \theta_1 \left[ \frac {C_\alpha \sqrt{2\theta_2h_0^2}}{\theta_1}+1+\frac{\theta_2h_0(h_0-1) }{\theta_1^2}\right]^{\frac{1}{h_0}},
\end{align}
where $\theta_i=\sum_{j=m+1}^{N} (\sigma_j)^i (i=1,2,3)$, $h_0=1-2\theta_1\theta_3/(3\theta_2^2)$, and $C_\alpha$ is the critical value of the normal distribution at test level of $\alpha$.

If the value of $Q_i$ is obviously beyond the control threshold $Q_\alpha$, there is a probability of $\alpha$ that the measurement error of one or more MMUs has drifted. By utilizing the $Q$ statistic, one can successfully identify the abnormal MMUs and the details of the method for correctly detecting all the abnormal MMUs will be described in the closely followed section.
\subsection{Robust Identification of the Abnormal MMUs} \label{sec: II-D}

{Based on the cyber-physics correlation analysis detailed in Section II-B, an inherent linear correlation exists among the magnetic field measurements from all MMUs. The principal component analysis method, highlighted in Section II-B as a robust technique for handling linear correlations, is well-suited for detecting changes in this correlation by monitoring the \(Q\) statistic against its threshold. Consequently, when measurement errors in one or more MMUs drift significantly, the proposed approach should be able to detect and identify them.}

In terms of practical realization, when $Q_i>Q_\alpha$, it can be highly guaranteed that the drift in measurement error occurs in the magnetic-array-type current sensor. To successfully locate all the MMUs with a drift in measurement error, a robust identification method is proposed, which is divided into the following two steps:
\begin{itemize}
    \item \emph{Step 1}: Select two normal MMUs without measurement error drifting. The existence of two normal MMUs is guaranteed by the truth that the number of the MMUs is sufficiently redundant, i.e., most of the MMUs work in a normal status. 
    
    To select two normal MMUs, all the MMUs are pairwise combined and the $Q$ statistic of each combination at each point of the test data is calculated. Then all the calculated results of the $Q$ statistic of each combination, over the whole data length of the test data, are summed. The combination with the minimum value of the above summation results of the $Q$ statistic is selected as the two normal MMUs.
    \item \emph{Step 2}: By combining the selected two normal MMUs with the remaining ones individually and comparing the value of the $Q$ statistic of the three-unit combination with the corresponding control threshold $Q_\alpha$, whether or not the MMU, added to the two selected normal MMUs, has a drift in its measurement error can be robustly determined. 
\end{itemize}

After all the abnormal MMUs are identified, the current measurement accuracy can be preserved by eliminating the measurement results of the abnormal MMUs when using (\ref{eq:1}). 
\section{Experimental Study}
\label{sec:03}
\subsection{Experimental Test Platform}

\begin{figure}[t!]
\centering
\includegraphics[width=\columnwidth]{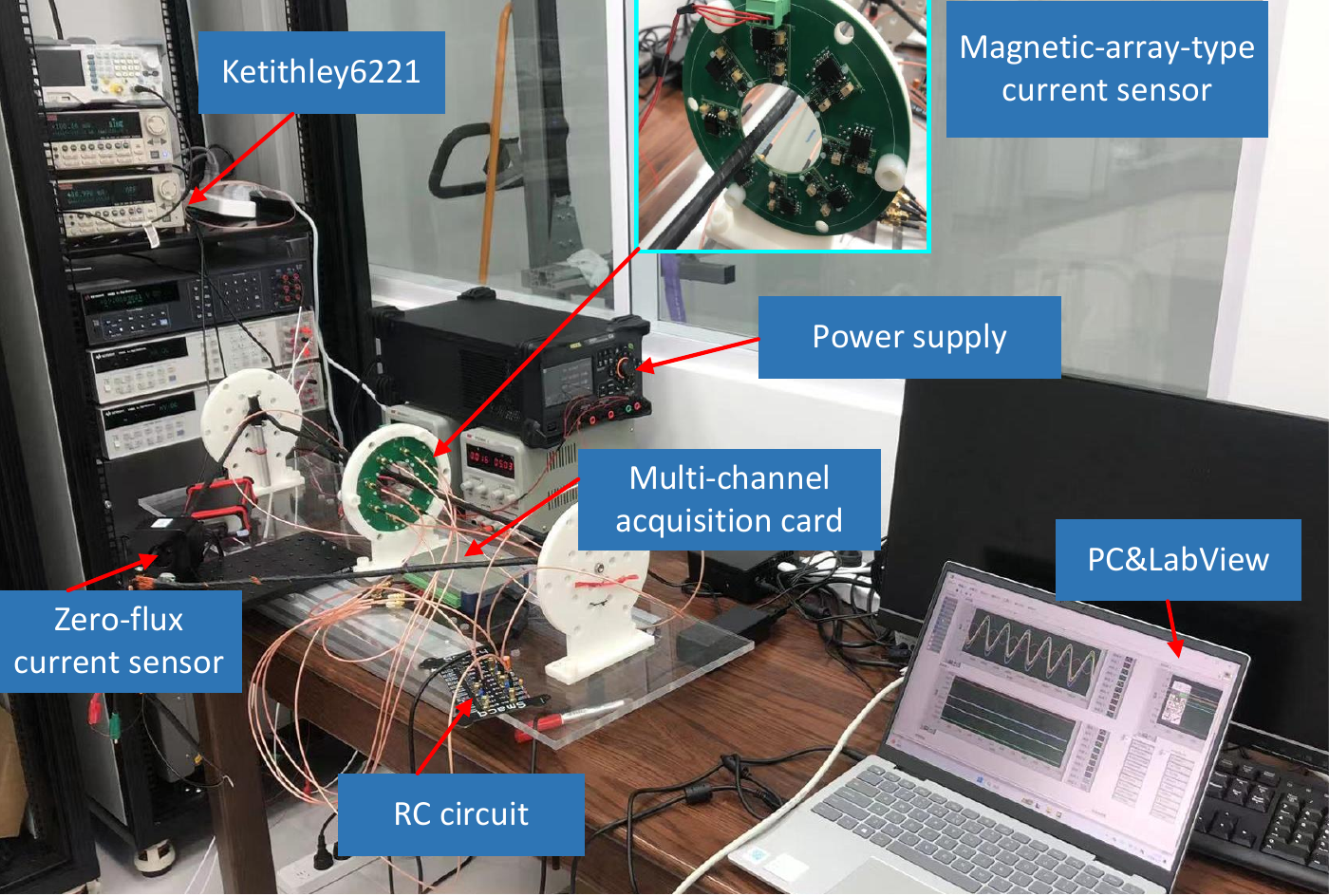}
\caption{A picture of the experimental setup.} 
\label{fig_tp}
\end{figure}

As shown in Fig. \ref{fig_tp}, a test platform is set up to verify the validation of the proposed self-healing approach. A magnetic-array-type current sensor is designed and fabricated, which is installed using a 3D printed structure. A conductor consisting of 100 turns of copper wire penetrates the current sensor. To eliminate interference from the ambient power frequency, an AC with a maximum amplitude of 105\,mA @60\,Hz, generated by a Keithley 6221 current source, is applied to the copper wire, allowing an AC with a maximum amplitude of 10.5\,A to be generated in the conductor. A highly accurate zero-flux current sensor, with a relative error of 0.01\% in amplitude and a phase error of $1\times 10^{-4}$\,rad @50\,Hz/60\,Hz, is used to measure the current flowing in the conductor as a reference. The output voltages of all the MMUs and the zero-flux current sensor are simultaneously sampled using a multiple-channel ADC with a sampling rate of 200\,kSa/s, 16-bit resolution, and an accuracy of 80\,ppm at its full range. 

\begin{figure}[!t]
\centering
\includegraphics[width=\columnwidth]{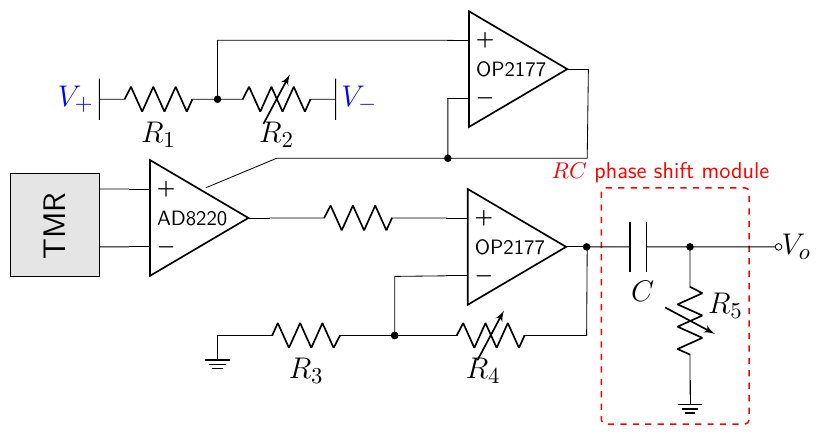}\\(a)\\
\includegraphics[width=0.8\columnwidth]{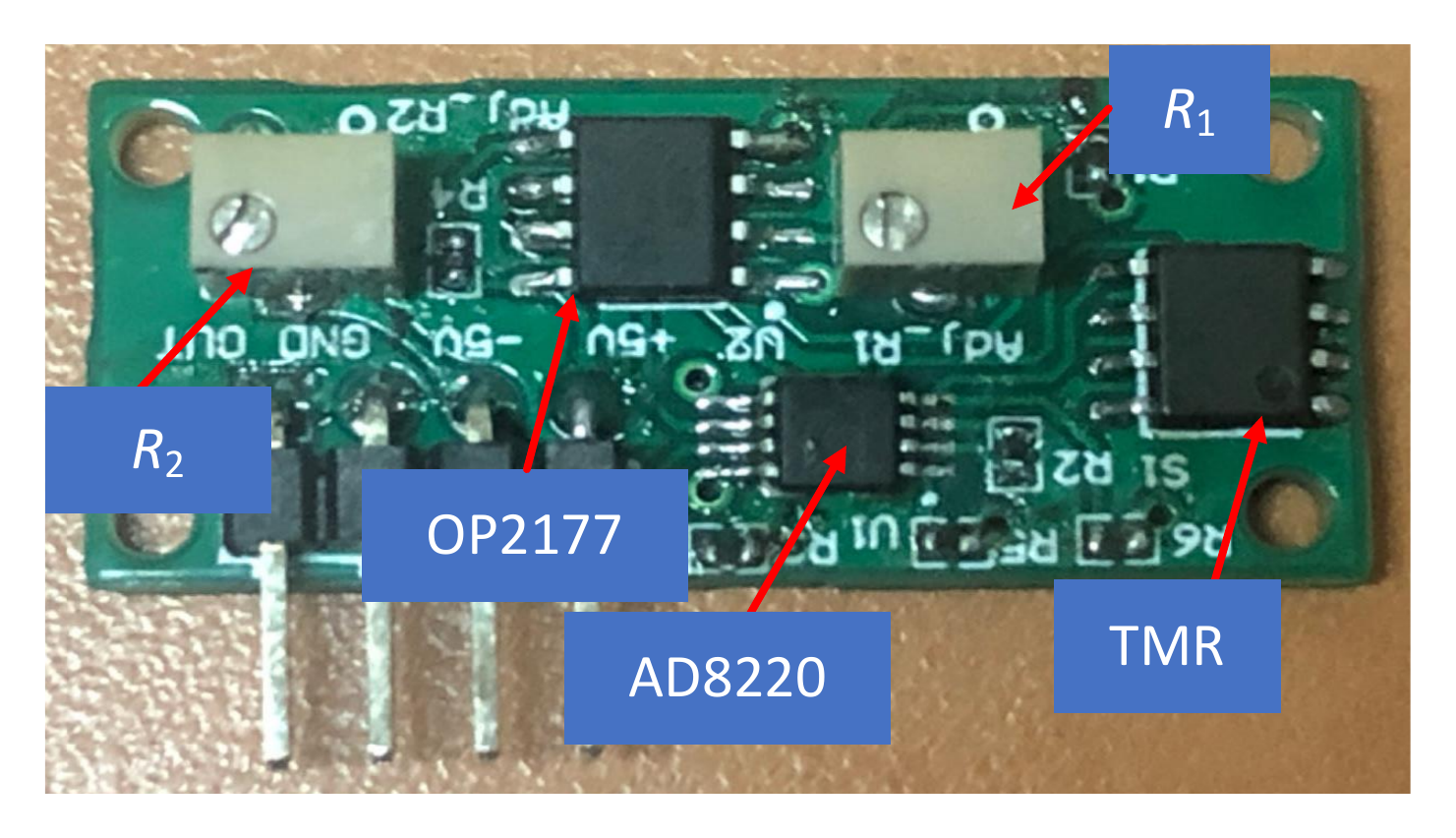}\\(b)
\caption{Designed MMU. (a) Design diagram of the MMU, (b) Practical electronic module of MMU.} 
\label{fig_du}
\end{figure}

The magnetic-array-type current sensor is composed of eight MMUs, as shown in Fig. \ref{fig_tp}. TMR element, produced by MultiDimension Technology Co., Ltd with a product NO. TMR2102, is used for magnetic field detection. TMR2102 has a linear range of $\pm$3\,mT and a sensitivity of 4.9\,mV/V/mT \cite{tmr2102} {with a 5\,V power supply voltage.} The circuit design of each MMU is shown in Fig. \ref{fig_du} (a). The TMR element converts the sensed magnetic field into output voltage, which is amplified using an instrument amplifier AD8220. The TMR element is configured in a Wheatstone full bridge structure using four magnetic field-sensitive units. Due to the non-ideality of the symmetry of the Wheatstone full bridge, a minor DC offset may exist in the output of the TMR element even though no magnetic field is applied. Thus, the potential reference of AD8220 is adjusted by tuning a resistive divider, consisting of $R_1$ and $R_2$, for DC offset compensation. Then the amplified voltage is further amplified by a second operation amplifier. 

\subsection{Calibration of the Magnetic-Array-Type Current Sensor}
The relationship between the output voltage of the $k$th MMU and the sensed magnetic field can be represented by
\begin{align}\label{eq:12}
\dot{B}_m^k &= g_ke^{i\varphi_k}\dot {U}_m^k,
\end{align}
where $\dot{U}_m^k$ is the output voltage phasor of the $k$th MMU, $g_k$ and $\varphi_k$, respectively, denotes the scale factor and phase difference between $\dot{B}_m^k$ and $\dot{U}_m^k$.  

Then, the expression (\ref{eq:1}) for current measurement can be converted to\cite{zhu2022event},
\begin{align} \label{eq:13}
\dot{I}_m &=\frac{1}{N}\sum_{k=1}^{N} \gamma_k{g_k}\dot{U}_m^ke^{i\varphi_k}\nonumber\\
&=\frac{1}{N}\sum_{k=1}^{N} \xi_k\dot{U}_m^ke^{i\varphi_k},
\end{align}
where $\xi_k$ denotes the scale factor between the measured current phasor $\dot{I}_m$ and the output voltage phasor $\dot{U}_m^k$ of the $k$th MMU.

\begin{figure*}[t]
\centering
\includegraphics[width=0.9\textwidth]{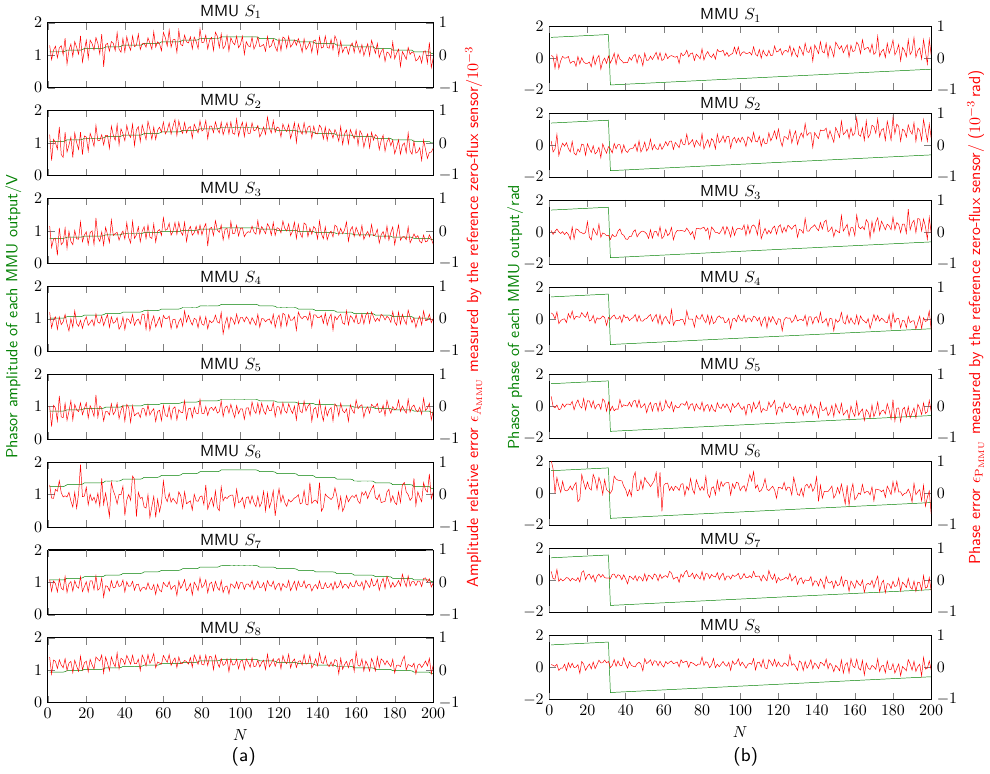}
\caption{Train data of each MMU. The amplitude of the excitation current (60\,Hz AC) is ramping in the form of a triangular wave. (a) shows the amplitude measurement results, where the left axis represents the phasor amplitude obtained from the MMU outputs, and the right axis is the amplitude relative error $\epsilon_{\rm{A_{MMU}}}$ referring to the measurements of the zero-flux current sensor. (b) presents the phase measurement results. The left axis is the phase of the phasor measured by each MMU and the right axis is the phase error $\epsilon_{\rm{P_{MMU}}}$ of each MMU compared to the reference value (zero-flux current sensor). Note that the transient of the phase curve is due to the phase range, $(-\frac{\pi}{2},\frac{\pi}{2}]$, defined by the \texttt{atan} function.} 
\label{fig_train_data}
\end{figure*}

The calibration of the magnetic-array-type current sensor is to determine the scale factor $\xi_k$ and the phase difference $\varphi_k$. By varying the amplitude of the current flowing in the conductor, the linear relationship between the amplitude of the current (measured by the precision zero-flux current sensor) and the output voltage of each MMU is obtained by performing linear fitting on the measured data, and the scale factor $\xi_k$ is obtained by calculating the slope of the fitted line. The nonlinearity (residual of the linear fit) of all the MMUs is less than $8\times10^{-4}$ and the phase difference of each MMU maintains almost constant over the range of the calibration current with a standard deviation less than $3\times 10^{-5}$\,rad.

\subsection{Experiments and Results}\label{sec:III-C} 
{Experiments are conducted to comprehensively validate the proposed self-healing approach. In these experiments, the measurement error drift of multiple MMUs is coupled by adjusting $R_4$ or $R_5$ of specific MMUs, as illustrated in Fig. \ref{fig_du} (a), or by changing the working temperature of a specific TMR element. The experimental procedures and results are detailed below.} 

\noindent\textit{1) Experiments with MMUs' Error Drifts Tuned Manually}
\label{sec:III-C 1)}

Experiments, where the measurement error drifts of multiple MMUs 
 are achieved by changing their scale factors or phase shifts by tuning $R_4$ or $R_5$ shown Fig. \ref{fig_du} (a), are first performed and the following steps are conducted:

\indent\textit{Step 1:\,Data Preparation.} After the calibration of the magnetic-array-type current sensor is accomplished, the output voltages of all the MMUs are simultaneously sampled with a multiple-channel ADC. Fluctuations of the input current are added to the test: the amplitude of the applied AC is set ramping in the form of a triangular wave. The amplitude and phase of each MMU are calculated and updated every 0.1\,s. A short period of the amplitude and phase measured in each MMU channel are respectively indicated in the left axis of Fig. \ref{fig_train_data} (a) and (b). 
 As can be seen from Fig. \ref{fig_train_data}, both the amplitude and the phase of all the MMUs vary, along with the fluctuation of the current to be measured, in the same trend, indicating a highly linear correlation between the measurement results of all the MMUs. Using the current phasor $\dot{I}$, measured by the zero-flux current sensor and the calibrated coefficients of all the MMUs, it can establish an experimental error monitoring mechanism. We define it as the reference phasor for each MMU, whose amplitude and phase are written as
\begin{subequations}
\label{eq:14}
\begin{align}
U_r^k &= \frac{I}{\xi_k}, \label{eq:14A}\\
\angle{\dot{U}_r^k} &= \angle\dot{I}-\varphi_k.\label{eq:14B}  
\end{align}
\end{subequations}

Note that the purpose of defining the reference phasor is to visualize the error change for each MMU, and the reference value is blind to MMUs. In such a way, we can compare the MMU errors identified by the data-driven proposal to these reference values.  

\begin{figure*}[t]
\centering
\includegraphics[width=0.9\textwidth]{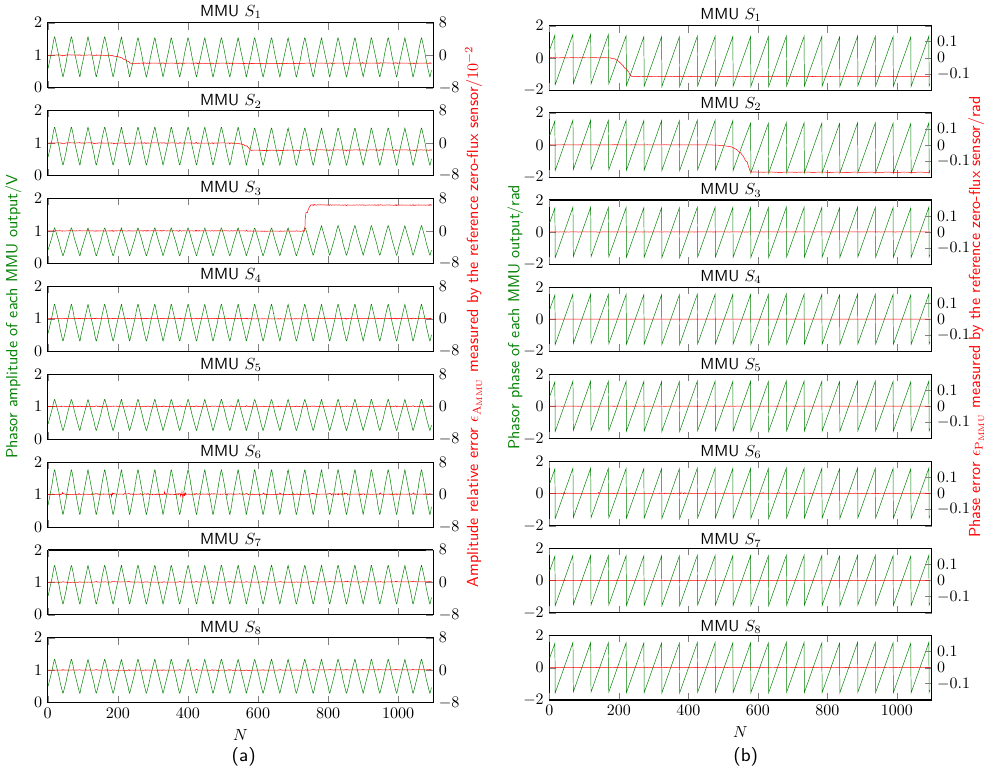}
\caption{Test data of each MMU. The amplitude of the excitation current (60\,Hz AC) is ramping in the form of a triangular wave. (a) shows the amplitude measurement, where the left axis represents the phasor amplitude obtained from the MMU outputs, and the right axis is the relative amplitude error $\epsilon_{\rm{A_{MMU}}}$ referring to the measurement of the zero-flux current sensor. (b) presents the phase measurement result. The left axis is the phase of the phasor measured by each MMU and the right axis is the phase error $\epsilon_{\rm{P_{MMU}}}$ of each MMU compared to the reference value (zero-flux current sensor). } 
\label{fig_test_data}
\end{figure*}

Based on the obtained reference values, the relative amplitude error and the phase error of each MMU can be obtained and they are, respectively, presented by the right axis of Fig. \ref{fig_train_data} (a) and (b). {As is seen from the amplitude relative error and phase error, the noise level of each MMU is not the same, due to the inconsistency of TMR elements used in the MMU fabrication. However, the amplitude relative error and the phase error for all 8 MMU channels are within the desired target, i.e. 0.1\% and $1\times10^{-3}$\,rad, respectively.}

With a good measurement accuracy for both amplitude and phase, the 20\,s measurement data presented in Fig. \ref{fig_train_data} are chosen as the training data for the input of the PCA. 
The amplitude data and the phase data of all the MMUs are respectively structured as two train-data matrices ${\bf{S}}_{L\times{N}}$. Using (\ref{eq:4})-(\ref{eq:6}), the principal component subspace $\bf{P}$ and the residual component subspace $\bf{V}$ for both signals can be separated. The residual component subspace $\bf{V}$ is kept unchanged and used over the whole online identification of the abnormal MMUs.

In reality, the online identification of abnormal MMUs can be performed over the long-term operation of the magnetic-array-type current sensor. However, in a laboratory experimental study, we can couple artificial errors into part of the MMU channels so that the proposal can be tested. Herein the amplitude errors are varied by changing the scale factor of the specific MMUs, i.e., adjusting the values of $R_4$ in Fig. \ref{fig_du} (a), while the amplitude error and phase error are simultaneously changed by adjusting the resistance $R_5$ in the phase shift module. Fig. \ref{fig_test_data} shows a set of test data for abnormal MMU identification. Similar to the train data, the excitation current also contains an amplitude modulation in the form of a triangular wave. The calculation update period, in this case, is 1\,s. As it can be seen from Fig. \ref{fig_test_data}, both the amplitude and phase measurements of all the MMUs are synchronized with the same trend along with the given triangular variation of the input current to be measured. In the error set, the amplitude relative errors and the phase errors of both MMUs $S_1$, $S_2$ gradually change in the direction of the negative error, and for MMU $S_3$, only its amplitude relative error is changed toward the positive direction. The remaining MMUs are unchanged and all working in their normal status. The red curves in Fig. \ref{fig_test_data} show the actual set errors referred to the zero-flux current sensor. These errors are at percent and sub-rad levels respectively for the amplitude and the phase, and visually these errors are fully immersed in the current ramping.

\begin{figure}[t]
\centering
\includegraphics[width=0.9\columnwidth]{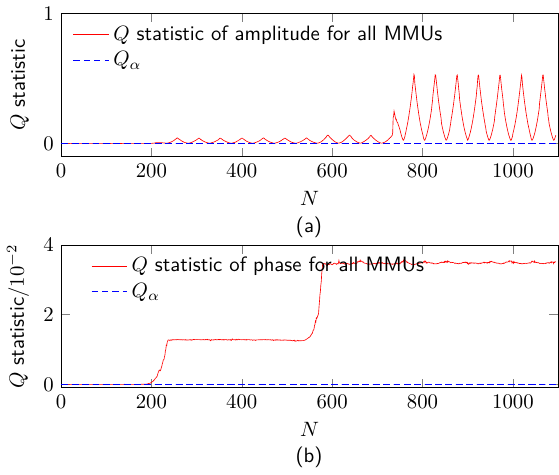}
\caption{(a) shows the $Q$ statistic of the amplitude of all MMU channels. (b) presents the $Q$ statistic of the phase for all MMUs. The horizontal axis is the time measurement sequence aligned with the test data as in Fig. \ref{fig_test_data}.}
\label{fig_q}
\end{figure} 

\begin{figure}[t]
\centering
\includegraphics[width=0.75\columnwidth]{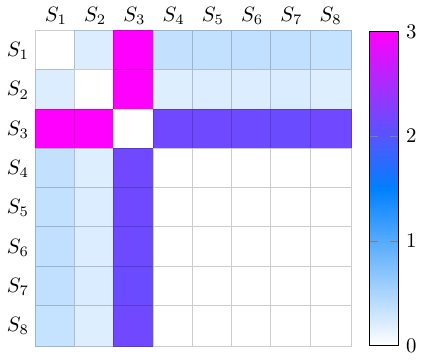}\\(a)\\
\includegraphics[width=0.75\columnwidth]{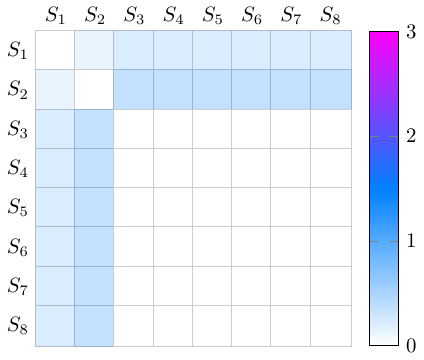}\\(b)
\caption{(a) Summation results of the $Q$ statistic for the amplitude of the voltage phasors of the pair-wise MMUs over the whole data length of the test data, (b) Summation results of the $Q$ statistic for the phase of the voltage phasors of the pair-wise MMUs over the whole data length of the test data.} 
\label{fig_si}
\end{figure}

\indent\textit{Step 2:\,Identification of Abnormal MMU(s).} The identification of three anomalous MMUs ($S_1$, $S_2$, and $S_3$) is accomplished through the proposed method delineated in Section \ref{sec: II-D}. The $Q$ statistic values, pertaining to both amplitude and phase across all MMUs, are acquired and visually represented in Fig. \ref{fig_q} (a) and (b). Analysis of the results reveals that the $Q$ statistics for both amplitude and phase surpass the predefined control threshold $Q_\alpha$ ($\alpha =0.99$). This observation implies discernible alterations in the measurement error for both amplitude and phase.

The first step in identifying abnormal MMUs is to find two robust normal MMU channels.
For this purpose, all the MMUs are pairwise combined and the $Q$ statistic of each combination is obtained using (\ref{eq:10}). The values of the $Q$ statistic of both the amplitude and the phase, considering all possible combinations of different channels, over the whole data length of the test data are, respectively, summed and the corresponding results are presented in Fig. \ref{fig_si}. Recall that in the test data, MMUs $S_1$ and $S_2$ have a measurement error drift in both amplitude and phase, and $S_3$ owns a measurement error drift only in amplitude. From Fig. \ref{fig_si}, it is obvious that the summed values of the $Q$ statistic, over the whole data length of the test data for the combinations with abnormal units, are much larger than those of those combinations with only normal units. From Fig. \ref{fig_si}, it can be evaluated that the MMUs $S_1$, $S_2$, and $S_3$ have relatively higher $Q$ statistic values and are likely to contain amplitude error drifts. Similarly, MMUs $S_1$ and $S_2$ have a relatively higher probability of having phase error drifts. 

It is possible to directly recognize the MMU channels that contain obvious amplitude error or phase error drifts. However, setting an appropriate threshold of the $Q$ statistic becomes tricky when multiple MMUs contain measurement error changes. On the opposite, it can always be highly convinced that the combinations with the lowest value in Fig. \ref{fig_si} are composed of only normal units, given that most of the MMUs are working in their normal status and there exist at least two normal MMUs. From Fig. \ref{fig_si}, MMUs $S_5$ and $S_7$ have the lowest $Q$ statistic and are determined as the normal reference MMUs. 
    
\begin{figure}[t]
\centering
\includegraphics[width=\columnwidth]{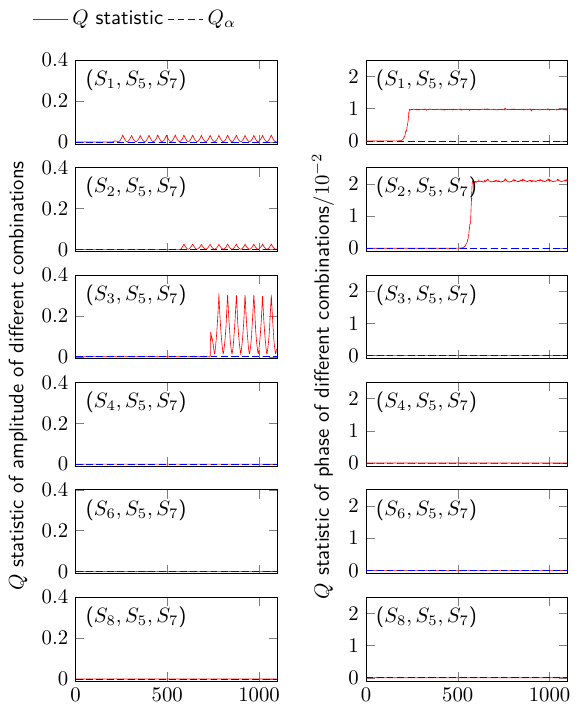}
\caption{Identification of normal and abnormal MMUs with measurement error drifts of multiple MMUs tuned manually. The left column subplots show the amplitude $Q$ statistic of combinations ($S_i$, $S_5$, $S_7$),  $i=1,2,3,4,6,8$ as a function of the measurement sequence. The right column subplots present the $Q$ statistic of the phase for the same combinations.} 
\label{fig_id}
\end{figure}

Further, the other MMUs are individually combined with the selected two normal reference MMUs ($S_5$ and $S_7$), and the $Q$ statistic of each three-unit combination is calculated. Fig. \ref{fig_id} shows the calculated results of the $Q$ statistic of three-unit combinations for both the amplitude and the phase: ($S_i, S_5, S_7$), $i=1,2,3,4,6,8$. It is obvious that the amplitude $Q$ statistic value of MMUs $S_1, S_2$ and $S_3$ exceeds their control threshold $Q_\alpha$, indicating these channels have an error drift of amplitude.  Similarly, the phase $Q$ statistic starts getting higher than the threshold value for MMUs $S_1$ and $S_2$, and hence the phase error change for these two channels are detected.  By comparing the $Q$ statistic value change over time to the error set shown in Fig. \ref{fig_test_data}, it can be seen that the $Q$ statistic can well identify when the MMU errors start to drift and how serious the error change is.  The $Q$ statistic value of the other three MMUs, i.e. $S_4$, $S_6$, and $S_8$, are comparable to the threshold value $Q_\alpha$ for both amplitude and phase, and therefore they are normal MMU channels.  

In summary, following the above steps, both the normal MMUs ($S_4$ to $S_8$) and the abnormal MMUs ($S_1$, $S_2$, $S_3$) can be successfully and robustly identified.

\begin{figure}[t]
\centering
\includegraphics[width=0.95\columnwidth]{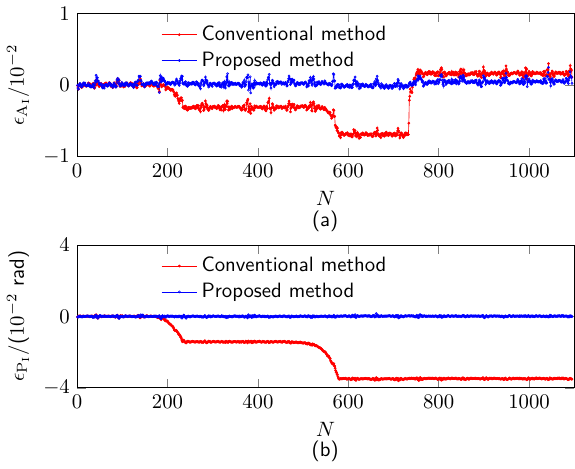}
\caption{A comparison of the current measurement results using conventional and proposed methods with measurement error drifts of multiple MMUs tuned manually. (a) shows the measurement results of the amplitude relative error with/without abnormal MMUs, and (b) shows the phase measurement error with/without abnormal MMUs.} 
\label{fig_current}
\end{figure}

\indent\textit{Step 3: Current Measurement Results.} After the abnormal MMUs are identified, the measurement results of only the normal MMUs are used to represent the current measurement result using (\ref{eq:13}). Fig. \ref{fig_current} (a) and (b) show the current measurement result of the proposed self-healing approach, where both the amplitude relative error $\epsilon_{\mathrm{A_I}}$ and the phase error $\epsilon_{\mathrm{P_I}}$ of the measured current phasor, referred to a zero-flux current sensor, are plotted.  In the same figure, the amplitude and phase measurement results of a conventional current measurement method, i.e. if the measurements of all the MMUs are used, are also presented for comparison.  

It can be seen from Fig. \ref{fig_current} that for a conventional approach, the current measurement result exhibits an error change closely correlated to the contribution of the measurement error drifts in abnormal MMUs. As it can be identified from Fig. \ref{fig_current} (a) the amplitude relative error continuously changes twice in the negative error direction, due to the error decrease of $S_1$ and $S_2$, and then increases towards a positive error direction arising from the error change of $S_3$. A maximum amplitude relative error of $-0.75\%$ is reached and the final error is $0.3\%$. From Fig. \ref{fig_current} (b), when all MMUs are used for current measurement, the phase error continuously decreases to negative error twice, due to the negative drift of the phase errors for $S_1$ and $S_2$, and finally reaches -0.036\,rad.
When the identified abnormal MMUs are eliminated as proposed, the amplitude relative error and the phase error of the current are respectively at a level of $0.2\%$ and $1.5\times10^{-3}$\,rad, which significantly demonstrates the validation of the proposed self-healing method for a magnetic-array-type current senor. 

In addition to the test result, from Fig. \ref{fig_current} (a), due to the amplitude error drifts in different directions ($S_1$ and $S_2$ changes towards negative error and $S_3$ varies in the direction of positive error), the final amplitude relative error is partially compensated. In this way, attributed to the random drifts of the measurement errors of the MMUs, the drift in measurement error of the magnetic-array-type current sensor can be mitigated to an extent, indicating the advantage of using the redundancy of the MMUs for the current measurement.

\begin{figure}[t]
\centering
\includegraphics[width=0.95\columnwidth]{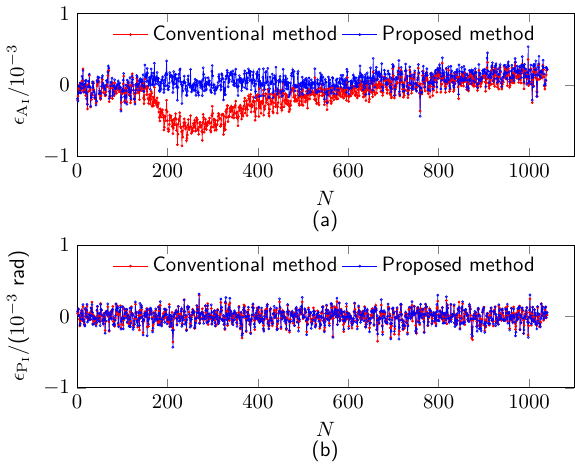}
\caption{A comparison of the current measurement results using conventional and proposed methods with measurement error drifts, caused by working temperature variation, of multiple MMUs. (a) shows the amplitude relative error $\epsilon_{\rm{A_I}}$ with/without abnormal MMUs, and (b) shows the phase error $\epsilon_{\rm{P_I}}$ with/without abnormal MMUs.} 
\label{fig_current_temp}
\end{figure}

{
\noindent\textit{2) Experiments with MMUs' Error Drifts Caused by Temperature Variation}\label{sec:III-C 2)}

During the long-term operation of magnetic-array-type current sensors, measurement errors of one or multiple MMUs may drift significantly due to non-uniform temperature distribution. To further validate the proposed self-healing approach, experiments involving variations in the working temperature of the MMUs are conducted.

Following the similar steps described in the above test example, the amplitude of the applied AC to be measured is modulated in a triangular form. The operating temperature of the MMU $S_6$ is varied by placing one end of a steel tweezer close to it, while the other end of the tweezer is heated by a soldering iron. The amplitude relative error $\epsilon_{\rm{A_{MMU}}}$ and phase error $\epsilon_{\rm{P_{MMU}}}$ of each MMU referring to a zero-flux current sensor are presented in Fig. \ref{fig_test_data_temp}. As shown in Fig. \ref{fig_test_data_temp}, the amplitude relative error of $S_6$ decreases significantly from nearly zero to approximately $-4\times10^{-3}$ and then returns to zero. This behavior is due to the rapid increase of working temperature for MMU $S_6$ when the soldering iron is turned on, followed by a gradual return to room temperature once the soldering iron is turned off. It is noteworthy that the amplitude relative error of two adjacent MMUs, $S_5$ and $S_7$, exhibits a similar trend to that of $S_6$, albeit with much smaller magnitudes, indicating they have detected the environmental temperature change caused by the heating of $S_6$. The phase error of each MMU remains unaffected by temperature variations.

\begin{figure}[t]
\centering
\includegraphics[width=0.95\columnwidth]{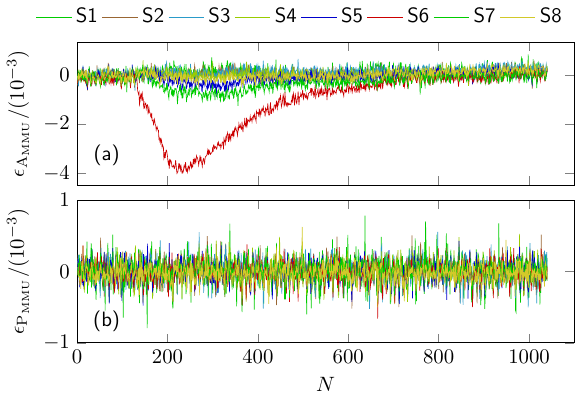}
\caption{Practical measurement errors of all MMUs caused by their working temperature variation. (a) shows the practical amplitude relative errors $\epsilon_{\rm{A_{MMU}}}$ of all MMUs, and (b) shows the practical phase errors $\epsilon_{\rm{P_{MMU}}}$ of all MMUs.} 
\label{fig_test_data_temp}
\end{figure}

According to \textit{step 2} in Section III-C 1), the abnormal MMUs can be identified. Initially, two normal reference MMUs (\(S_3\) and \(S_4\)) are recognized. The other MMUs are then individually combined with these two reference MMUs, and both the amplitude and phase of the \(Q\) statistic for the three-unit combinations are calculated: \((S_i, S_3, S_4)\), \(i=1,2,5,6,7,8\). Fig. \ref{fig_id_temp} presents the calculated \(Q\) statistic results and the corresponding control value \(Q_\alpha\). It can be seen from Fig. \ref{fig_id_temp} that the amplitude \(Q\) statistic of the three-unit combination \((S_6, S_3, S_4)\) increases significantly above its control value, indicating a measurement error drift in \(S_6\), before decreasing back to the control value. This trend aligns with the amplitude relative error of \(S_6\) shown in Fig. \ref{fig_test_data_temp} (a). Therefore, the \(Q\) statistic can effectively reveal the measurement error drift of each MMU. Furthermore, the amplitude \(Q\) statistic for the three-unit combinations \((S_5, S_3, S_4)\) and \((S_7, S_3, S_4)\) also shows a similar variation trend, slightly exceeding the \(Q\) statistic control value, indicating that the measurement error drifts of \(S_5\) and \(S_7\) are detectable but much smaller compared to \(S_6\). This observation is consistent with the practical error drifts shown in Fig. \ref{fig_test_data_temp} (a). From Fig. \ref{fig_id_temp}, the phase \(Q\) statistic of each three-unit combination remains close to its control value, which means there is no considerable phase error drift for all MMUs.

\begin{figure}[t]
\centering
\includegraphics[width=0.95\columnwidth]{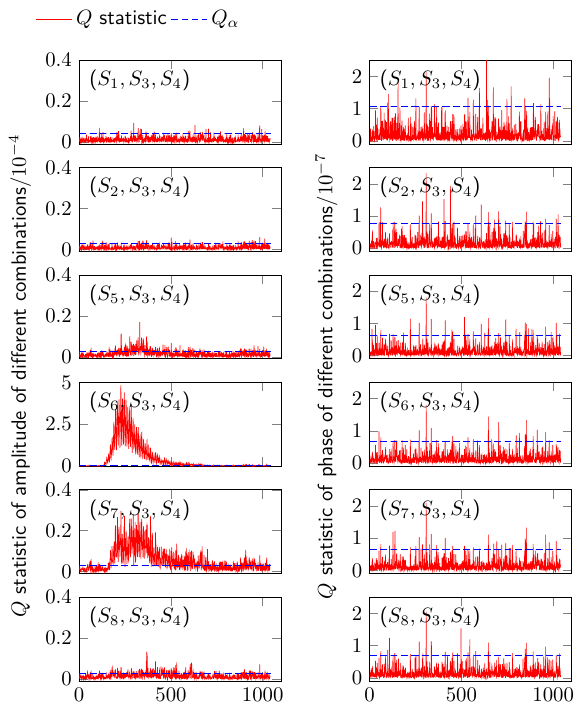}
\caption{Identification of normal and abnormal MMUs with measurement error drifts, caused by working temperature variation, of multiple MMUs. The left column subplots show the amplitude $Q$ statistic of combinations ($S_i$, $S_3$, $S_4$), $i=1,2,5,6,7,8$ as a function of the measurement sequence. The right column subplots present the $Q$ statistic of the phase for the same combinations.} 
\label{fig_id_temp}
\end{figure}

From Fig. \ref{fig_id_temp}, $S_5$, $S_6$, and $S_7$ can be identified as the MMUs with abnormal measurement error drifts. Fig. \ref{fig_current_temp} shows the current measurement results using the proposed self-healing approach, which only utilizes the measurement results from the normal MMUs to represent the current measurement results as described in (\ref{eq:13}). Both the amplitude relative error $\epsilon_{\rm{A_I}}$ and phase error $\epsilon_{\rm{P_I}}$ of the measured current phasor, referred to the current measurement results of a zero-flux current sensor, are plotted. In the same figure, the current measurement results of the conventional method, which uses the measurement results of all MMUs, are also presented for comparison.

In Fig. \ref{fig_current_temp} (a), the measurement result of $\epsilon_{\rm{A_I}}$ using the conventional approach exhibits a variation trend similar to the amplitude error of the identified abnormal MMUs ($S_5$, $S_6$, and $S_7$). Specifically, it decreases from nearly zero to approximately -0.08\% and then returns to nearly zero, closely correlated with the temperature change applied. In contrast, the proposed self-healing approach maintains an amplitude relative error $\epsilon_{\rm{A_I}}$ close to zero, regardless of temperature changes. This demonstrates that the proposed self-healing approach is capable of detecting measurement error drifts for $\epsilon_{\rm{A_I}}$ less than 0.08\%, which is well below the typical error threshold (no better than 0.2\%) of a metrology-level magnetic-array-type current sensor used in smart grid. As can be seen from Fig. \ref{fig_current_temp} (b), the phase error of the current measurement results by both the conventional method and the proposed method in this paper is essentially zero because the phase error of the MMUs is insensitive to temperature change.}

\section{Conclusion}
\label{sec:04}
{Magnetic-array-type current sensors offer significant potential for non-contact current measurement across various voltage levels. A bottleneck for such kind of senor is that during long-term operation, its measurement accuracy may degrade due to error drifts (caused by environmental factors for example) of one or more MMUs. Traditional offline calibration methods are time-consuming, labor-intensive, and involve a heavy workload. Thus, how to online monitor the error status of MMUs and hence the measurement accuracy of magnetic-array-type current sensors during long-term operation becomes a worth exploring issue. 
This paper proposes a data-driven, self-healing approach to enhance the long-term stability and reliability of these sensors. In the proposal, the inherent linear correlation between the measurement results of all MMUs is derived from cyber-physical correlation analysis, and the PCA, known for its proficiency in controlling linear correlation analysis, is then used for identifying MMUs error status change. Further, a robust criterion based on \(Q\) statistic is proposed to identify abnormal MMUs. After the abnormal MMUs are identified, the current represent algorithm is updated by using only the measurement results of unbiased MMUs, and therefore the measurement accuracy of the sensor is preserved.

Experimental studies are performed to verify the proposed idea. Two tests are carried out using different error generation mechanisms. In the first experiment, measurement errors of a few chosen MMUs are modified by adjusting their scale factors and phase shifts. The amplitude relative error and the phase error reach $-0.75\%$ and -0.036\,rad with the conventional approach, while the errors are reduced to within $0.2\%$ and $1.5\times10^{-3}$\,rad after the proposed method is applied. In the second test, error drifts due to the thermal change are considered. The heating of a soldering iron is coupled to MMUs. A maximum measurement error of 0.08\% of the current sensor, mainly due to a 0.4\% amplitude error just in one MMU, is observed. By identifying and eliminating the abnormal MMUs using the proposal, the 0.08\% measurement error drift can be removed. Both tests demonstrate the effectiveness of the method proposed in this paper.
         
Except for current measurement applications, the proposed self-healing approach holds the potential for extension to various array-type sensors, provided there is adequate redundancy. Examples include electric field sensors for non-contact voltage measurement \cite{li2023knowledge}, gravitational field sensors for gravity measurement \cite{xu2023development}, mass sensors using magnetic suspension balance for contactless mass measurement \cite{wang2023contactless}, acoustic sensors for acoustic source localization \cite{oliveri2024acoustic}, etc.}

\section*{Acknowledgement}
The authors would like to thank industrial colleagues from the China Southern Power Grid and the State Grid Corporation of China for valuable discussions. This work was supported by the National Key Research
and Development Program under Grant 2022YFF0708600. 

\begin{thebibliography}{10}

\bibitem{li2020research}
Z.~Li, W.~Jiang, A.~Abu-Siada, \textit{et al}, ``Research on a composite voltage and current measurement device for HVDC networks,'' \emph{IEEE Transactions on Industrial Electronics}, vol.~68, no.~9, pp. 8930--8941, 2020.

\bibitem{sun2021ltcc}
P.~Sun, X.~Cui, S.~Huang, \textit{et al}, ``LTCC based current sensor for silicon carbide power module integration,'' \emph{IEEE Transactions on Power Electronics}, vol.~37, no.~2, pp. 1605--1614, 2021.

\bibitem{wu2023improved}
Y.~Wu, Z.~Wei, Y.~Yang, and P.~Zhang, ``Improved common-mode leakage current measurement method for insulation condition monitoring in distribution grids,'' \emph{IEEE Transactions on Industrial Electronics}, vol.~71, no.~5, pp. 5307--531, 2023.

\bibitem{ziegler2009current}
S.~Ziegler, R.~C. Woodward, H.~H.-C. Iu, and L.~J. Borle, ``Current sensing techniques: A review,'' \emph{IEEE Sensors Journal}, vol.~9, no.~4, pp. 354--376, 2009.

\bibitem{kyriazis2023modeling}
G.~A. Kyriazis and R.~S. Ribeiro, ``Modeling the transimpedance amplitude and phase of wideband cage-type current shunts,'' \emph{IEEE Transactions on Instrumentation and Measurement}, vol.~72, pp. 1502407, 2023.

\bibitem{locci2000digital}
N.~Locci and C.~Muscas, ``A digital compensation method for improving current transformer accuracy,'' \emph{IEEE Transactions on Power Delivery}, vol.~15, no.~4, pp. 1104--1109, 2000.

\bibitem{jiang2023fiber}
R.~Jiang and Y.~Zheng, ``Fiber-optic current sensing based on reflective polarization-bias-added structure,'' \emph{IEEE Sensors Journal}, vol.~24, no.~1, pp. 292--302, 2023.

\bibitem{khan2021magnetic}
M.~A. Khan, J.~Sun, B.~Li, \textit{et al}, ``Magnetic sensors-a review and recent technologies,'' \emph{Engineering Research Express}, vol.~3, no.~2, pp. 022005, 2021.

\bibitem{moghe2012novel}
R.~Moghe, F.~Lambert, and D.~Divan, ``A novel low-cost smart current sensor for utility conductors,'' \emph{IEEE Transactions on Smart Grid}, vol.~3, no.~2, pp. 653--663, 2012.

\bibitem{li2022contactless}
P.~Li, B.~Tian, L.~Li, \textit{et al}, ``A contactless current sensor based on TMR chips,'' \emph{IEEE Transactions on Instrumentation and Measurement}, vol.~71, pp. 9511711, 2022.

\bibitem{liu2022coreless}
X.~Liu, W.~He, P.~Guo, and Z.~Xu, ``A coreless current probe for multicore cables,'' \emph{IEEE Sensors Journal}, vol.~22, no.~20, pp. 19282--19292, 2022.

\bibitem{liu2023enhanced}
J.~Liu, M.~Guan, Y.~Xu, \emph{et~al}, ``Enhanced limit-of-detection of current sensor based on tunneling magnetoresistive effect with multi-chips differential design,'' \emph{IEEE Transactions on Instrumentation and Measurement}, vol.~72, pp. 2008009, 2023.

\bibitem{liu2022semi}
X.~Liu, W.~He, P.~Guo, \textit{et al}, ``Semi-contactless power measurement method for single-phase enclosed two-wire residential entrance lines,'' \emph{IEEE Transactions on Instrumentation and Measurement}, vol.~71, pp. 9508216, 2022.

\bibitem{khawaja2017estimating}
A.~H. Khawaja and Q.~Huang, ``Estimating sag and wind-induced motion of overhead power lines with current and magnetic-flux density measurements,'' \emph{IEEE Transactions on Instrumentation and Measurement}, vol.~66, no.~5, pp. 897--909, 2017.

\bibitem{kazim2019fault}
M.~Kazim, A.~H. Khawaja, U.~Zabit, and Q.~Huang, ``Fault detection and localization for overhead 11-kV distribution lines with magnetic measurements,'' \emph{IEEE Transactions on Instrumentation and Measurement}, vol.~69, no.~5, pp. 2028--2038, 2019.

\bibitem{zhu2017energization}
K.~Zhu, W.~K. Lee, and P.~W. Pong, ``Energization-status identification of three-phase three-core shielded distribution power cables based on non-destructive magnetic field sensing,'' \emph{IEEE Sensors Journal}, vol.~17, no.~22, pp. 7405--7417, 2017.

\bibitem{ibrahim2020design}
M.~E. Ibrahim and A.~M. Abd-Elhady, ``Design and modeling of a two-winding Rogowski coil sensor for measuring three-phase currents of a motor fed through a three-core cable,'' \emph{IEEE Sensors Journal}, vol.~21, no.~6, pp. 8289--8296, 2020.

\bibitem{lee2018superpositioning}
S.~Lee, Y.~Ahn, T.~Kim, \textit{et al}, ``A superpositioning technique for accurate open-type current sensing in three-phase electrical switchboards,'' \emph{IEEE Sensors Journal}, vol.~18, no.~22, pp. 9297--9304, 2018.


\bibitem{ma2019impact}
X.~Ma, Y.~Guo, X.~Chen, \textit{et al}, ``Impact of coreless current transformer position on current measurement,'' \emph{IEEE Transactions on Instrumentation and Measurement}, vol.~68, no.~10, pp. 3801--3809, 2019.

\bibitem{zhang2019current}
H.~Zhang, F.~Li, H.~Guo, \textit{et al}, ``Current measurement with 3-D coreless TMR sensor array for inclined conductor,'' \emph{IEEE Sensors Journal}, vol.~19, no.~16, pp. 6684--6690, 2019.

\bibitem{chen2022intelligent}
K.-L. Chen, ``Intelligent contactless current measurement for overhead transmission lines,'' \emph{IEEE Transactions on Smart Grid}, vol.~13, no.~4, pp. 3028--3037, 2022.

\bibitem{zhu2022event}
Q.~Zhu, G.~Geng, and Q.~Jiang, ``Event-driven non-invasive multi-core cable current monitoring based on sensor array,'' \emph{IEEE Transactions on Power Delivery}, vol.~38, no.~3, pp. 1548--1557,  2022.

\bibitem{liu2021nonintrusive}
X. Liu, W. He, Y. Zhao, \textit{et al}, "Nonintrusive current sensing for multicore cables considering inclination with magnetic field measurement," \emph{IEEE Transactions on Instrumentation and Measurement}, vol. 70, pp. 9513314, 2021.


\bibitem{huang2014magnetics}
Q.~Huang, Y.~Song, X.~Sun, \textit{et al}, ``Magnetics in smart grid,'' \emph{IEEE Transactions on Magnetics}, vol.~50, no.~7, pp. 0900107, 2014.

\bibitem{duan2023operational}
J.~Duan, Q.~Tang, J.~Ma, and W.~Yao, ``Operational status evaluation of smart electricity meters using Gaussian process regression with optimized-ARD kernel,'' \emph{IEEE Transactions on Industrial Informatics}, vol.~20, no.~2, pp. 1272--1282, 2023.

\bibitem{zhang2022online}
Y.~Zhang, C.~Zhang, H.~Li, and Q.~Chen, ``An online detection method for capacitor voltage transformer with excessive measurement error based on multi-source heterogeneous data fusion,'' \emph{Measurement}, vol. 187, pp. 110262, 2022.

\bibitem{chu2023automatic}
R.~Chu, L.~Chik, J.~Chan, \textit{et al}, ``Automatic meter error detection with a data-driven approach,'' \emph{Engineering Applications of Artificial Intelligence}, vol. 123, pp. 106466, 2023.

\bibitem{itzke2018influence}
A.~Itzke, R.~Weiss, and R.~Weigel, ``Influence of the conductor position on a circular array of Hall sensors for current measurement,'' \emph{IEEE Transactions on Industrial Electronics}, vol.~66, no.~1, pp. 580--585, 2018.

{\bibitem{yu2018circular}
H.~Yu, Z.~Qian, L.~Huayi, and Q.~Jiaqi, "Circular array of magnetic sensors for current measurement analysis for error caused by position of conductor,"\emph{Sensors}, vol. 18, no. 2, pp. 578, 2018.}

\bibitem{greenacre2022principal}
C.~Labrín and U. Francisco. ``Principal component analysis,''In R for Political Data Science, Chapman and Hall/CRC, pp. 375-393, 2020.

{\bibitem{tmr2102} TMR2102 Datasheets, MultiDimension Technol. Co., Jiangsu, China.}

{\bibitem{li2023knowledge}
H.~Li, C.~Ma, C.~Zhang, \textit{et al}, "A knowledge-based cooperative co-evolutionary algorithm for non-contact voltage measurement," \emph{IEEE Transactions on Emerging Topics in Computational Intelligence}, vol.~8, no.~2, pp. 1142--1155, 2024.
}

{\bibitem{xu2023development}
F.~Xu, M.~Song, C.~Chen, and X.~Xu, "Development of a compact magnetic suspension balance," \emph{IEEE Transactions on Instrumentation and Measurement}, vol. 72, pp. 1001209, 2023.}

{\bibitem{wang2023contactless}
Y.~Wang, L.~Jiang, Z.~Chen, \textit{et al}, "Contactless weighing method based on deep learning and acoustic levitation,"\emph{Measurement Science and Technology}, vol. 35, no.~5, pp. 056005, 2024.}

{\bibitem{oliveri2024acoustic}
M.~Olivieri, A.~Bastine, M.~Pezzoli, \textit{et al}, "Acoustic imaging with circular microphone array: A new approach for sound field analysis," \emph{IEEE/ACM Transactions on Audio, and Language Processing}, vol. 32, pp. 1750--1761, 2024.}

\end{thebibliography}

\end{document}